\documentclass[twocolumn,showpacs,preprintnumbers,amssymb,superscriptaddress]{revtex4-1}
\usepackage{graphicx}
\usepackage{dcolumn}
\usepackage{bm}
\usepackage {longtable}
\usepackage{braket}
\usepackage{epstopdf}
\usepackage{amsmath}
\input{Qcircuit.tex}

\begin{document}

\bibliographystyle{apsrev4-1}
\preprint{}

\title{A cold-atoms based processor for deterministic quantum computation with one qubit in intractably large Hilbert spaces}

\author{C. W. Mansell}
\author{S. Bergamini}
\affiliation{The Open University, Walton Hall, Milton Keynes, MK7 6AA} \date{\today}

\begin{abstract}

We propose the use of Rydberg interactions and ensembles of cold atoms in mixed state for the implementation of a
protocol for deterministic quantum computation with one quantum bit (DQC1)  that can be readily operated in high
dimensional Hilbert spaces. We propose an experimental test for the scalability of the protocol and to study the
physics of discord. Furthermore we develop a scheme to add control to non-trivial unitaries that will enable the study
of many-body physics with ensembles in mixed states.
\end{abstract}


\maketitle

\section{INTRODUCTION}

At present, no single feature of the quantum world has been identified as the source of the computational enhancement,
efficiency and speed-up of quantum protocols. Whilst entanglement is widely recognised as a key resource in quantum
technology \cite{revdisc}, an exponential advantage over classical computing can be achieved without it
\cite{PhysRevX.1.021022} in the presence of non-classical correlations (discord). Experiments using few photonic qubits
\cite{ExperimentalQuantum2008bpLanyon} have shown that some computational tasks that are classically intractable can be
efficiently solved even with no entanglement.  The dynamics of entanglement and discord differ considerably, with
entanglement being extremely fragile towards decoherence (even undergoing entanglement "sudden death"
\cite{Almeida27042007}) and discord being much more robust \cite{revdisc}. Since decoherence is a major hurdle to the
development of quantum technologies \cite{PhysRevA.80.024103, PhysRevA.81.052318}, the investigation of protocols that
are more robust against it is a promising route for progressing the field.

In the past years, there has been outstanding progress in the demonstration of quantum algorithms based on pure states
with a limited number of qubits. However scalability remains an issue, mainly because of decoherence. In pure-states
quantum computation (QC) this problem can possibly be  solved by error correction. Nevertheless, scaling up to a
significant number of qubits and being able to perform a classically intractable calculation has been impossible so
far.

Deterministic quantum computation with one qubit (DQC1) is a non-universal model  based on mixed states that can
exponentially speed up some computational tasks for which no efficient classical algorithms are known. DQC1 protocols
present a remarkable advantage with respect to standard QC protocols, in that it requires only a single qubit with
coherence to perform large scale quantum computation, whilst its power scales up with a number of qubits in mixed
state. It is therefore in principle more readily scalable, provided a suitable system for the implementation is
developed. Although it has been shown that this scheme contains little to no entanglement
\cite{EntanglementAnd2005aDatta}, non-classical correlations are present in the output state of the DQC1 which can be
quantified in terms of quantum discord \cite{QuantumDiscord2008aDatta}. Discord has been shown to be a valuable
resource for  specific computational tasks and for being extremely robust towards decoherence
\cite{PhysRevA.80.024103}, which is the stumbling block in developing quantum technologies \cite{naturedatta},
\cite{revdisc}. To date, successful experiments based on DQC1 have evaluated the normalised trace of a two-by-two
unitary matrix \cite{ExperimentalQuantum2008bpLanyon} and performed the approximation to the Jones polynomial with a
system of four qubits \cite{ExperimentalApproximation2009gPassante}, thus demonstrating the ground principle of mixed
state computation. However, these experiments were performed with photons and nuclear magnetic resonance respectively,
with limited scalability so far. Eventually, like for pure states quantum computation,  the protocol is useful only if
it can be scaled up and run over a significant number of qubits. Therefore DQC1 needs to be tested and operated in
large Hilbert spaces, so it is vital to benchmark it in a system that allows to reach this regime.

We propose a new scheme to  investigate experimentally the physics of DQC1 and discord in many-atom ensembles for a
specific algorithm that performs the normalized trace estimation \cite{PowerOf1998eKnill}. Cold ensembles in
micron-sized dipole traps can contain a few to hundreds of atoms and we find that the protocol under study is robust
enough to be operated both in small and large ensembles. We demonstrate that applying the protocol on an ensemble of
100 atoms will successfully evaluate the normalized trace of a  $2^{100}$-by-$2^{100}$  matrix. Finding the normalised
trace of this matrix is equivalent to adding up about $10^{30}$ numbers, which is a task that is classically
intractable for non-trivial matrices. More importantly, the scheme we developed can quantify geometric discord in the
system \cite{QuantumDiscordAs2012, QuantumDiscordIn2012yYao, arxivref1, OperationalMeaningOf2012sAdhikari}, therefore
it allows a systematic study of the computational power of discord. Besides providing an experimental test of the
protocol in high-dimensional Hilbert space and ultimately a test of the scalability and resilience of mixed-state
computation,  this work proposes an application of DQC1 to measure the mean-field interaction strength in a large
ensemble of many interacting particles. Finally we provide a general scheme to extend the protocol to different
controlled unitaries, such as those encounterd in many-body physics, quantum thermodynamics and quantum metrology
\cite{PhysRevLett.110.230601, PhysRevLett.110.230602}.

\begin{figure}
\includegraphics[scale=0.33]{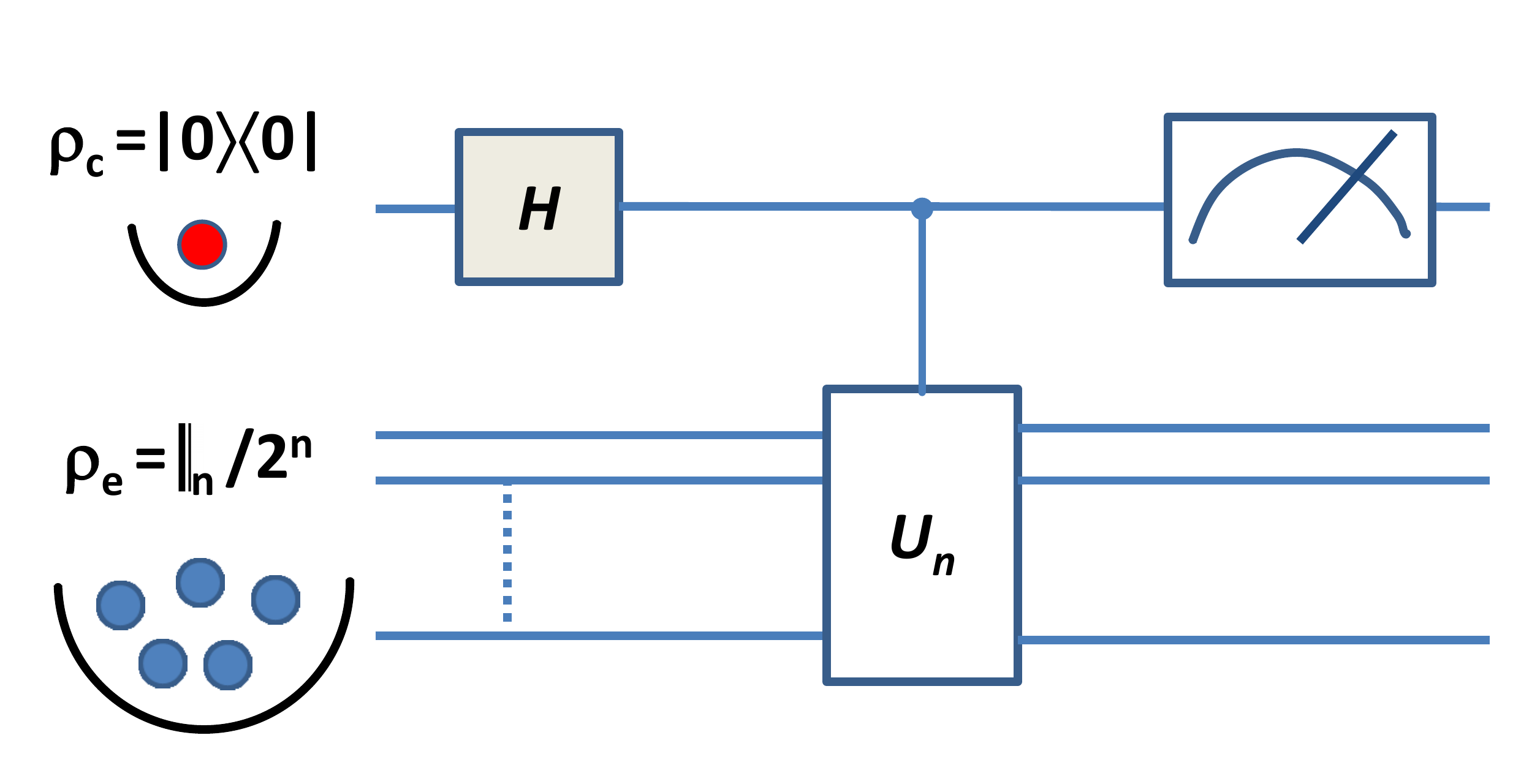}
\caption{
\label{dqc1}(Color online).
Circuit model of the DQC1 algorithm. The control atom and the ensemble are optically trapped at a
distance and individually optically addressed. The control atom is prepared in a pure state
$\ket{0}\bra{0}$ whilst the ensemble atoms are prepared in the maximally mixed state.}
\end{figure}

\section{DQC1} Fig. \ref{dqc1} describes the DQC1 algorithm: the input state consists of a single control qubit, whose
purity can be varied, prepared in the state $\ket{0}\bra{0}$ and a register of n qubits which are in the maximally
mixed state $I_{n}/2^n$. After a Hadamard operation on the single qubit, a controlled unitary $ U_{n}$ is performed on
the n-qubits mixed state.

This has the effect of encoding the normalized trace of the unitary operation into the single qubit coherences, and the
output state of the control qubit can be written as:
\begin{equation}
\rho_{C_{out}}= \frac{1}{2}\left(
                                \begin{array}{cc}
                                1 & \frac{Tr[U^{\dag}_{n}]}{2^n} \\
                                \frac{Tr[U_{n}]}{2^n} & 1
                              \end{array}
                              \right)
\end{equation}

The trace of the unitary $ U_{n}$ can then be retrieved by measurement of the expectation values of the Pauli operators
(X and Y) on the single qubit, as $\langle \emph{X} \rangle = Re[Tr(U_n)]/2^{n}$ and $\langle \emph{Y} \rangle =
-Im[Tr(U_n)]/2^{n}$.

\section{DQC1 with atoms}
In the scheme we propose, the control qubit can be stored either in the ground states of a single atom or in an
ensemble of strongly interacting atoms, using techniques that have been recently proposed to prepare and control
mesoqubits \cite{mesoqubit}. For clarity we will refer to the control qubit as a single atom qubit, but an extension of
the protocol to a mesoqubit is straightforward.  The single atom qubit is used as the control for a unitary operation,
enabled by Rydberg-Rydberg interactions, on a register of $n$ qubits encoded in an ensemble of atoms, as shown in Fig.
\ref{dqc1}.

\begin{figure}
\includegraphics[scale=0.33]{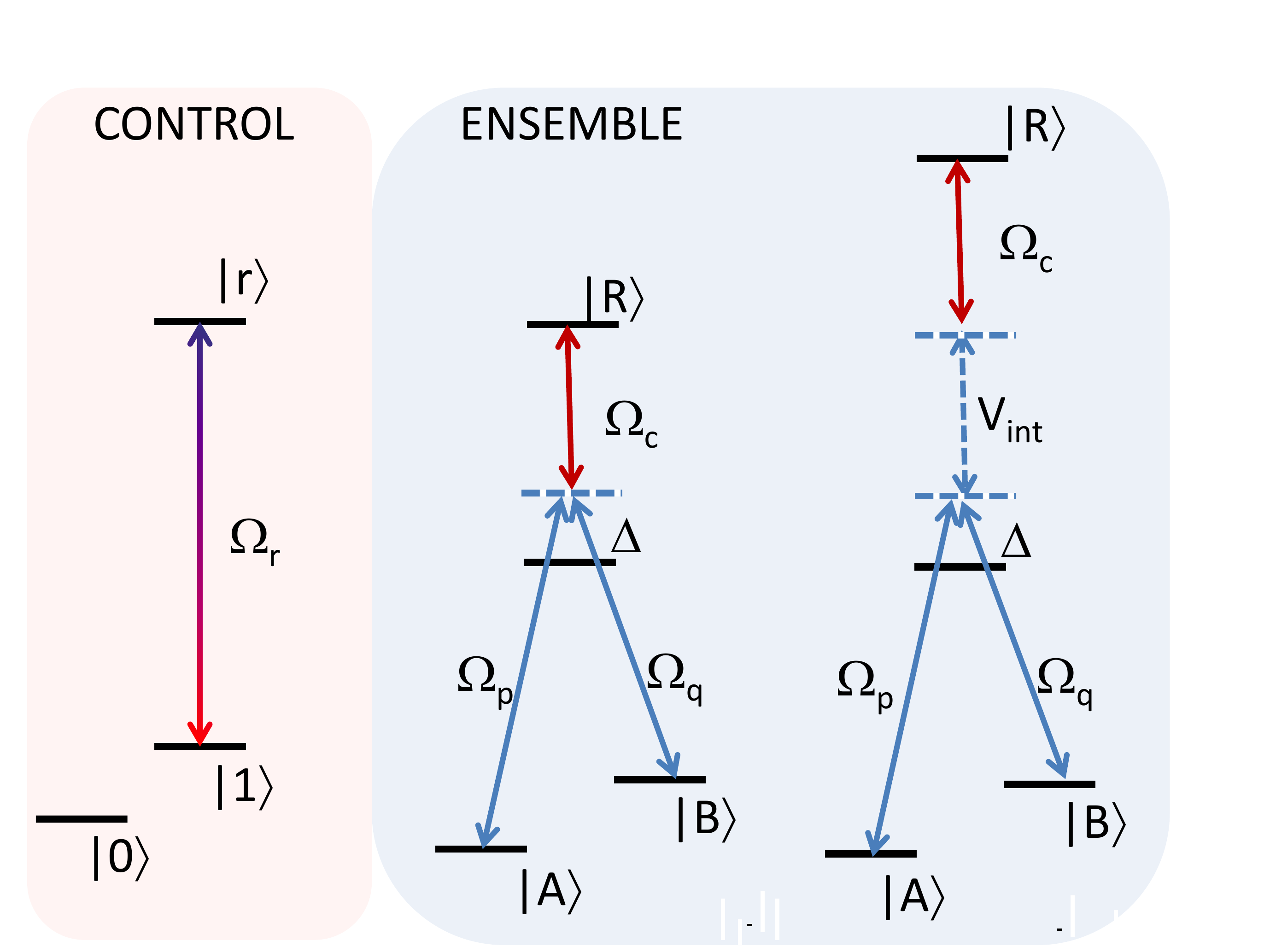}
\caption{
\label{bench}(Color online).
Optical scheme to implement the controlled off-resonant Raman rotation. The control qubit is encoded in the states $\ket{0}$ and $\ket{1}$ of a single atom.
State $\ket{1}$ is coupled to a Rydberg state via $\Omega_{r}$. Each of the  \textit{n}  qubits in the ensemble is encoded  in the states
$\ket{A}$ and $\ket{B}$ which are coupled by a 2-photon scheme  similar to \cite{MesoscopicRydberg2009mMuller}. A beam coupling the
intermediate state to the Rydberg state is added so that the EIT condition is fulfilled and the interaction with $\Omega_{p}$ and  $\Omega_{q}$
 is inhibited (left panel). However the coupling of  the control atom to Rydberg state can activate an additional shift that removes the
 condition for EIT, so that off-resonant Raman transfer is activated (far-right panel).}
\end{figure}

The control qubit and the ensemble are stored in two separate micron-sized dipole traps that are individually
addressable \cite{LaserTrapping2004sBergamini} \cite{MesoscopicRydberg2009mMuller}.  In the case under study the qubit
is encoded in the two ground state hyperfine levels of $^{87}$Rb (in Fig. \ref{bench} represented by $\ket{0}$ and
$\ket{1}$). The ensemble qubits are first encoded in the same hyperfine ground states of $^{87}$Rb (in Fig. \ref{bench}
represented by $\ket{A}$ and $\ket{B}$), and the ensemble is subsequently prepared in a highly mixed state. The single
qubit acts as a control atom over the target ensemble via excitation to Rydberg state and we use a laser excitation
scheme developed in \cite{MesoscopicRydberg2009mMuller} based on electromagnetically induced transparency (EIT) and
shown in Fig. \ref{bench}.

We performed numerical calculations to study the feasibility of the experimental implementation of the  protocol to
benchmark this method for a specific choice of the unitary and for different number of atoms in the ensemble.

 \subsection{Initialization:} The control qubit is first prepared  via optical pumping  in state $\ket{1}$.
 A $\pi$-pulse (Hadamard rotation via stimulated Raman transition) is then performed to initialize the qubit in $\ket{+}=\frac{1}{\sqrt{2}}(\ket{0}+\ket{1})$
 (Table I). This can be achieved with a fidelity $> 99.9\%$, as discussed in \cite{RydbergState2011mSaffman}. The control qubit is therefore prepared in
 a superposition of states $\ket{0}$ and $\ket{1}$, and, in general,  its  purity can be varied.
 \begin{table}
 \begin{tabular}{|c|c|c|c|c|c|c|}
   \hline
& \multicolumn{1}{c}{Initialization}&
 \multicolumn{3}{|c|}{Processing}&
 \multicolumn{2}{c|}{Measure}\\
   \hline

C & $\ket{+}=\frac{1}{\sqrt{2}}(\ket{0}+\ket{1})$& Ryd $\pi$  &  & Ryd $\pi$ & X(Y)  & fluo\\

   \hline
  E &  $I_{n}/2^{n}$  & &$CU_{n}$  & &   &  \\
  \hline

 \end{tabular}
 \\
 \caption{
 Summary of the sequence of operations on the control and ensemble qubits to perform the DQC1 protocol. After the initialization stage the qubit is prepared in $\ket{+}=\frac{1}{\sqrt{2}}(\ket{0}+\ket{1})$
 and the ensemble in a maximally mixed state. The processing stage sandwiches a controlled unitary between two $\pi$-pulses
 (the first couples state $\ket{1}$ to a Rydberg state
 and the second returns back to the ground state), so that the control  qubit acquires some Rydberg character necessary to operate the
 controlled unitary and it is then returned to its original state.
 Fluorescence  measurements are performed on the populations of states $\ket{0}$ and $\ket{1}$ after an X-(Y) rotation.}
 \end{table}
 Similarly, the ensemble state is obtained by first preparing  a $50/50$ weighted superposition $\ket{+}_n=\frac{1}{\sqrt{n}}(\ket{0} + \ket{1})^{\otimes n}$
 by applying the Hadamard gate to the whole ensemble containing \textit{n} atoms.
 To introduce the ``mixedness"  we propose a method adapted from a scheme developed for ions
 \cite{ExperimentalMultiparticle2010jtBarreiro}: following the preparation of state $\ket{+}_n$,
  one of the two ground states is  coupled to the intermediate state  $\ket{P}$ so that optical pumping  exposes the ensemble to decoherence and a
  mixed state is prepared. We operate the optical pumping over a stretched state, so that there is no loss of population and the mixed state can be
  prepared with optimum efficiency.

\begin{figure}[h!] \centering
\begin{minipage}[t]{0.5\textwidth}
\includegraphics[width=\textwidth]{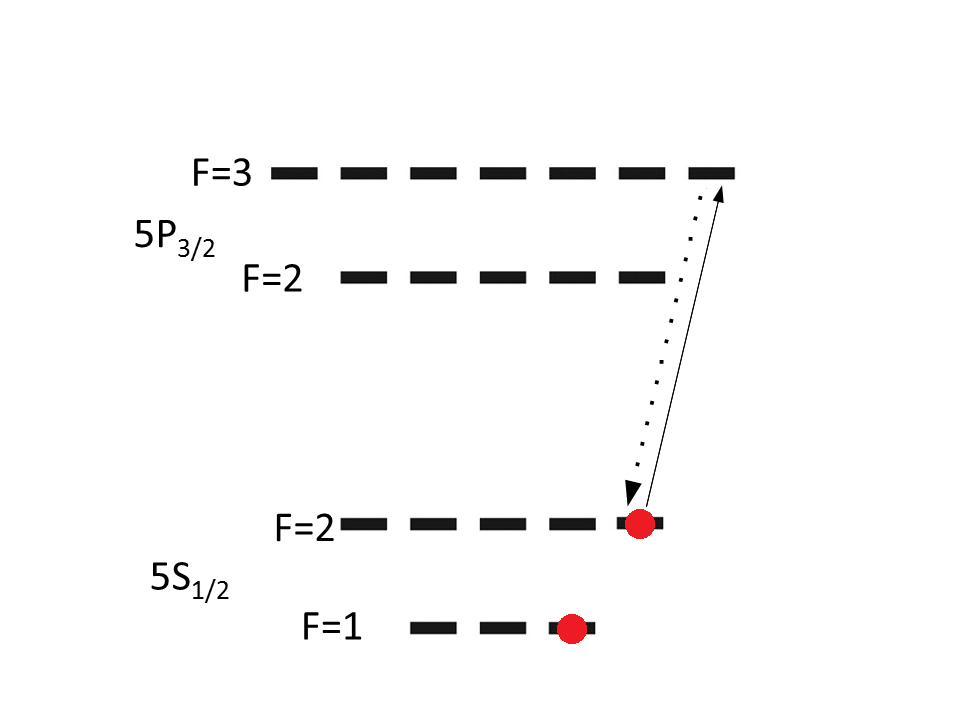}
\end{minipage}
\label{purity}
\caption{States $\ket{0}$ and $\ket{1}$ are $5S_{1/2}$, $F=1$, $m_F=1$ and $5S_{1/2}$, $F=2$, $m_F=2$ respectively. Following the preparation of state $\ket{+}_n$,
  state $\ket{1}$ is  coupled to the intermediate state  $5P_{3/2}$, $F=3$, $M_F=3$  via $\sigma$- polarised light, so that optical pumping  exposes the ensemble to decoherence and a
  mixed state is prepared. This method is used both to prepare the ensemble in a maximally mixed state and to vary the purity of the control qubit.}
\end{figure}

  The purity of the control qubit and the mixedness of the ensemble are controlled using the same level scheme
  described in figure \ref{purity}

 \subsection{Processing:}
 The DQC1 protocol relies on a controlled unitary  performed on an ensemble of atoms prepared in a highly mixed state.
 To benchmark the protocol  we choose to apply a controlled non-resonant Raman rotation to the ensemble atoms qubits. This is done exploiting a scheme
 similar to the one developed in \cite{MesoscopicRydberg2009mMuller} for CNOT gates.  We find that the protocol, described in Fig. \ref{bench},
  works very efficiently with high fidelity for any controlled-rotations.

 The processing stage begins with a $\pi$ pulse applied to the control atom so that the coupling between state $\ket{1}$  $\ket{r}$ is activated, as
 shown in the Table I.
An off-resonant Raman pulse is then applied to the ensemble atoms  to performs rotations of the ensemble qubits
corresponding to different angles in the Bloch sphere.

We performed  simulations of this scheme  by numerically solving the time dependent Schrodinger equation for a 4-level
atomic system in the presence of finite Rydberg blockade and taking into account decay from the intermediate state.  We
find that, for high fidelity operation for both small and large n,  the following conditions have to be fulfilled: i)
The Raman detuning $\Delta$ has to be much larger than  the inverse of the decay rate of the intermediate state, to
make sure that spontaneous decays is highly suppressed, ii) the lifetime of the Rydberg state chosen for the control
atom has to be much larger than the operation time of the controlled Raman and iii) $\Omega_p$,$\Omega_q  \ll \Omega_c$
to ensure that when the control atom is not in the Rydberg state, the EIT condition is met and there is little unwanted
coupling of the ensemble atoms to the light. This is in agreement with \cite{MesoscopicRydberg2009mMuller}, where this
laser scheme was used to perform controlled logic on ensemble atoms.

\begin{figure}
\includegraphics[scale=0.45]{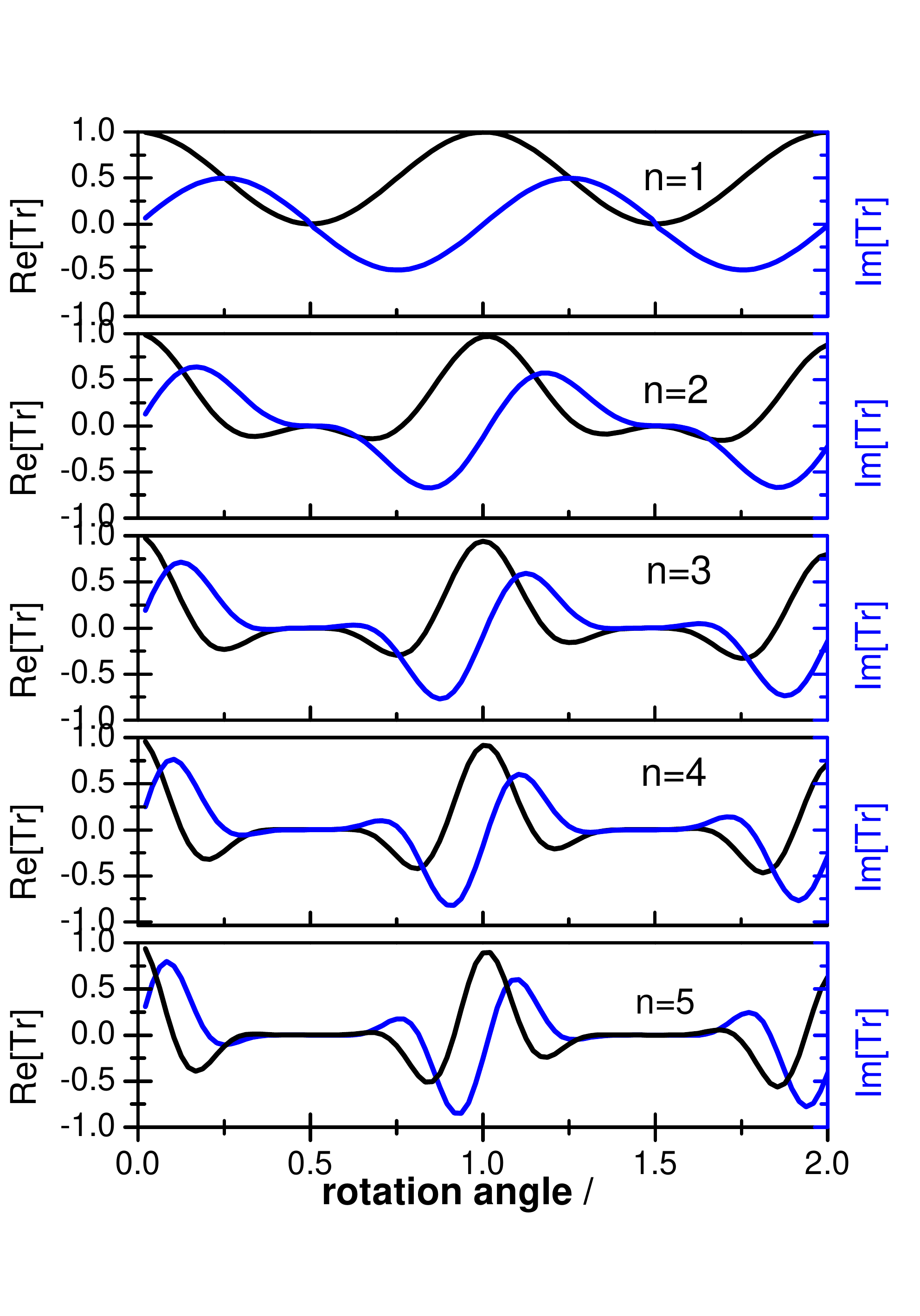}
\caption{
\label{results}(Color online).
 Results of the numerical estimate of the real (black) and imaginary (blue) parts of the normalized trace for
 $\Omega_{p} = \Omega_{q} =2 \pi \times 70 $ MHz,
  $ \Delta = 2 \pi \times 1200$ MHz  from the intermediate state. The decay rate $2 \pi \times 6$ MHz from the intermediate state
  is also taken into account. $\Omega_c$ is
   chosen to be $2 \pi \times 700$ MHz. $\ket{R} = 63 S$ and $\ket{r} = 64 S$ for Rubidium 87 that,
   for  a separation between the  traps of $1.7 \mu$m,
 provide an interaction strength of $15$ GHz. We take into account the decay from the intermediate state.}
\end{figure}

We choose $\ket{R} = 63 S$ and $\ket{r} = 64 S$ for Rubidium 87 that, for  a separation between the  traps of $1.7
\mu$m, provide an interaction strength  in excess of $15$ GHz. The Raman beams both have Rabi frequency $\Omega_{p} =
\Omega_{q} = 2 \pi \times 70 $ MHz and detuning $ \Delta = 2 \pi \times 1200$ MHz   from the intermediate state.
$\Omega_c$ is  chosen to be $2 \pi \times 700$ MHz. This coupling Rabi frequency can be obtained with commercially
available intermediate power laser sources focused down to waists of tens of micrometers. With these parameters, the
EIT-induced blocking of the Raman transfer works with a fidelity of more than $99.8\%$
\cite{MesoscopicRydberg2009mMuller}. We numerically calculate the evolution of the system after a pulsed Raman rotation
of different duration (i.e. corresponding to a different angle in the Bloch sphere) and we retrieve the X(Y)
expectation values. We  find that  the real and imaginary  part of the trace of the unitary acting on the ensemble of
atoms take the form shown  in Fig. \ref{results} for different number of atoms in the ensemble. The slight damping in
time of the amplitudes of the peaks reflects a dynamical phase shift, extensively discussed in
\cite{MesoscopicRydberg2009mMuller}.

\subsection{Measure of the trace and geometric discord} At the end of the protocol, the measurement of the state of the
control qubit will allow us to retrieve  the real and imaginary part of the trace of the unitary respectively. This is
done by statistical measurements of the populations of $\ket{0}$ and $\ket{1}$ following an X-(Y-)rotation.
X-(Y-)rotations can be performed with very high fidelity so that they negligibly affect the fidelity of the measurement
result \cite{ramanrot}. To measure the expectation value with an accuracy $\epsilon$ requires the number of runs to be
$NR \sim 1/\epsilon^2$, as shown in \cite{EntanglementAnd2005aDatta}. It is important to note that the number of runs
necessary for a set accuracy does not depend on the number of qubits in the ensemble. Furthermore, the populations are
measured via fluorescence imaging, that  also suffers  for limited efficiency and significant error rate, particularly
when working with single atoms. In order to achieve better than a 10\% accuracy requires averages over 400 runs.

\begin{figure}
\includegraphics[scale=0.9]{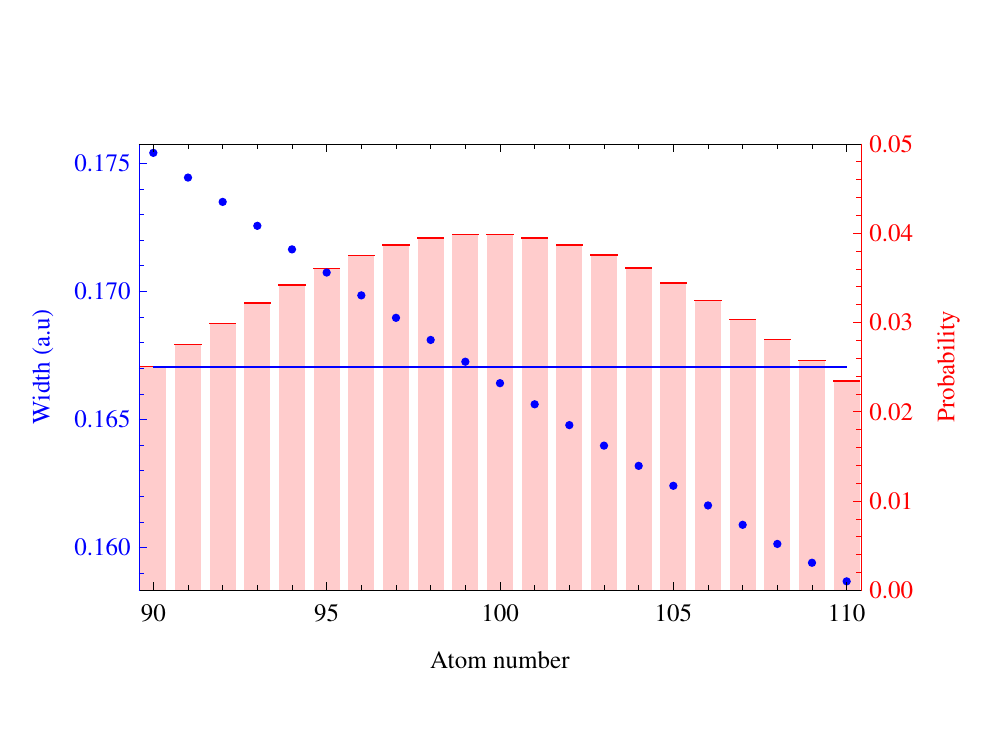}
\caption{
\label{analytical}(Color online).
 Width of the features in the real part of the normalized trace for different atom number, according to a Poissonian distribution of atom number
 with average $\overline{n}$=100.
   Plotted are the probability of loading the trap with a given number of atoms (red) and
  the width of the peaks in the trace versus atom number (blue). The solid blue line represents the weighted average of the
  of the widths, i.e. the result of the measurements.}
\end{figure}

 It needs to be
pointed out that both the control atom and the ensemble atoms are randomly loaded in small size dipole traps
\cite{LaserTrapping2004sBergamini}. The trap can be operated in controlled regimes, so that a single atom can be loaded
with probability $80\%$ \cite{singleatoms, LaserTrapping2004sBergamini} and it is possible to conditionally start the
experiment once an atom is loaded. The ensemble is typically loaded with a Poisson-distributed number of atoms around
an average value $\overline{n}$. At small n, we can force the number of atoms in the trap to be exactly \textit{n} for
every run of the experiment by post-selection and retrieve the traces in Fig. \ref{results} with small uncertainties.
But whilst this is reasonable at small \textit{n}, it would reduce the efficiency of the protocol at high \textit{n}.
We find, however, that for high atom number benchmarking and test for discord can be done by locking of the average
number of atoms (which can be tuned by parameters such as trap depth and density of the reservoir).  We have estimated
the uncertainties in the value of the trace measured arising from the fluctuations in atoms number in the ensemble from
run to run. We assume a Poissonian distribution of atom number with average number 100,
  as in Fig. \ref{analytical}.
 The height of the peaks are found to be insensitive to the atom number for the parameters
 chosen in this work. However, as shown in Fig. \ref{analytical}, the width of the features detected in the trace narrows with increased atom number,
 leading to an uncertainty in the value of the trace measured. For an average atom number $\overline{n}=100$ Poissonian variations from run to run lead to an uncertainty
of less than $5\%$ at all points (it is negligible at the peak, it is maximum at the position of fast variation).

Other sources of uncertainties  related to Rydberg interactions within the ensemble do not affect significantly the
fidelity of the protocol, provided a suitable choice of $\Omega_{p,q,c}$ is made \cite{MesoscopicRydberg2009mMuller}.

Finally, the geometric discord can be quantified by simply performing the controlled-unitary twice in a row in the DQC1
implementation \cite{MeasuringGeometric2012gPassante}, and this measure has the same accuracy discussed for the trace
estimation.

\section{Implementing non-trivial unitaries} Besides testing the scalability of DQC1 it will be interesting to extend the DQC1 protocol to the
implementation of \textit{non-trivial} controlled unitary. In general U can be time-evolution operator of some physical
system and the ability to enable control qubits extends the range of operability of the protocol. We therefore
identified a scheme for $X_a$ \cite{AddingControlTo2011x-qZhou} gate on the ensemble that allows us to add the control
to a range of unitaries of interest.

\subsection{The X$_a$ gate}

Any quantum operation (e.g. a unitary evolution) can be made to depend on the state of a control qubit, using the
general results presented in Ref. \cite{AddingControlTo2011x-qZhou}. This work demonstrates the equivalence between a
controlled unitary and a sequence of a controlled-X$_{\textit{a}}$ gate followed by the quantum operation and the same
CX$_{\textit{a}}$ gate afterwards, as shown in Fig. \ref{UsefulnessOfControlled_Xa}. This result simplifies the task of
finding interesting controlled unitaries that could be implemented in the Rdyberg-DQC1 experiment into the task of
designing a cold atom version of the controlled-X$_{\textit{a}}$ gate.

\begin{figure}[h!]
\centering
\Qcircuit @C=.5em @R=0em @!R {
   \push{\rule{7em}{0em}}   & \ctrl{1}  & \qw      &\ctrl{1}  & \qw & \push{\rule{0.3em}{0em}=\rule{0.3em}{0em}} & \ctrl{1} & \qw \\
   \push{\rule{7em}{0em}}   & \gate{Xa} & \gate{U} &\gate{Xa} & \qw & \push{\rule{0.9em}{0em} \rule{0.9em}{0em}} & \gate{U} & \qw
}
\caption{Circuit identity showing how two controlled-X$_{\textit{a}}$ gates can be used to implement an arbitrary unitary, U,
in a controlled way.}
\label{UsefulnessOfControlled_Xa}

\end{figure}

\begin{figure}
\centering
\includegraphics[scale=0.5]{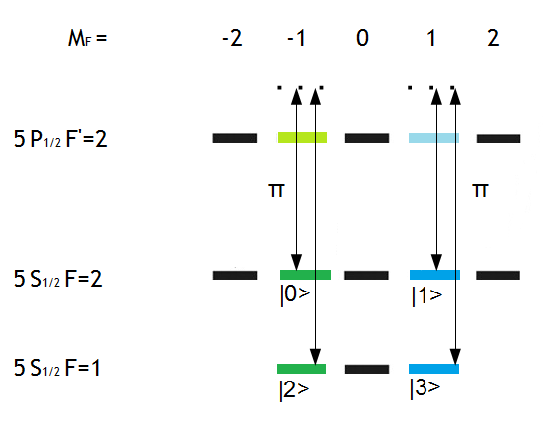}
\caption{(Color Online).
Transfer between $|0\rangle = |5S_{1/2},F=2,m_{F}=-1\rangle$ and $|2\rangle = |5S_{1/2} F=1, mF=-1\rangle$ (shown in dark green),
is provided by a controlled off-resonant Raman transition using linearly polarised light via $5P_{1/2} F^{\prime}=2, mF^{\prime}=-1$ (shown in light green).
Transfer between $|1\rangle = |5S_{1/2},F=2,m_{F}=1\rangle$ and $|3\rangle = |5S_{1/2},F=1,m_{F}=1\rangle$ (shown in dark blue),
is provided by a controlled off-resonant Raman transition using linearly polarised light via $5P_{1/2} F^{\prime}=2, mF^{\prime}=1$ (shown in light blue).
When the transfer pulses are $\pi$-pulses, an X$_{\textit{a}}$ gate is implemented. A controlled-X$_{\textit{a}}$ gate
can be performed by adding a coupling laser so that EIT occurs conditionally depending on the state of a control atom,
using the scheme in \ref{bench} and different qubit states. A magnetic field has to be added to lift the degeneracy of
the Zeeman states (not represented in figure).} \label{fig c-Xa}
\end{figure}
 To explain in more detail, the X$_{\textit{a}}$ gate operates on a
four-dimensional Hilbert space spanned by the qubit states, $|0\rangle$ and $|1\rangle$, and two auxiliary states,
$|2\rangle$ and $|3\rangle$. The auxiliary states are chosen so that they are not acted upon by the quantum operations
that act on the qubit states. The truth table for the X$_{\textit{a}}$ gate reads:

\begin{equation}
\begin{array}{cc} Xa |0\rangle = |2\rangle & Xa |1\rangle = |3\rangle \\ Xa |2\rangle = |0\rangle & Xa |3\rangle = |1\rangle   \end{array}
\end{equation}

Here we propose a scheme to perform a controlled X$_{\textit{a}}$ exploiting Rydberg blockaded Raman transitions,
 where a controlled off-resonant Raman scheme enables the transfer of atoms
 from the qubit basis to the auxiliary one, conditional on the state of the control qubit.
In figure \ref{fig c-Xa} the atomic level and the simplified light diagram is summarized and explained. More details on
the complete Raman scheme can be found in \cite{PhysRevA.72.022347}. A magnetic field has to be added to lift the
degeneracy of the Zeeman states (not represented in figure).  Our choice of Zeeman states is governed by the
consideration that pairs of states $|F, M_{F}\rangle$ and $|F+1,-M_{F}\rangle$ experience the same linear Zeeman shifts
(see references \cite{mSaffman2005AnalysisOfAQuantum} \cite{eBrion2007MultilevelEnsembles},
\cite{eBrion2008ErrorCorrectionInEnsemble}).

\subsection{Many-body physics}

In the paragraph above we have presented a method to add control to any unitary that operates on a given basis via the
$X_a$ gate. The task of finding a  range of interesting controlled unitaries that can be implemented using this method
is therefore simplified.

The study of the time evolution of many-body interacting systems is certainly one of the key drivers for quantum
simulators. The DQC1 protocol with atoms would  implement a variety of operations and can be used to explore the
physics of interacting system. As an example,  by coupling the qubit basis to a Rydberg state we can  switch on
\textit{interactions} within the ensemble. Efficient coupling to Rydberg states can be obtained using the schemes
described in \cite{PhysRevA.84.023413}. The requirement of leaving the auxiliary states unaffected is quite easily met
because of the large separation between the hyperfine levels of the ground state (6.8 GHz). It can be shown that, in
the regime of strong interactions (i.e. Rydberg-Rydberg interaction strength much larger than Rabi couplings) a measure
of the normalised trace of the time-evolution operator at different times will retrieve an average value over the
ensemble for the interaction strength.

Finally, this implementation with cold atoms is extremely versatile. The `target' atoms can be arranged in arrays of
dipole traps \cite{JOSAB, michal} where each site is individually addressable, so that different unitary operations can
be performed on different qubits.

 This work will motivate the design of protocols to solve a range of problems
computationally hard, like finding the ground state of the 2- or 3-dimensional Ising model with a local transverse
field with interactions beyond nearest neighbours \cite{fBarahona1982OnTheComputational} or studying  the unitaries
involved in collisions of Rydberg polaritons \cite{jStanojevic2012GeneratingNon-Gaussian}.

\section{Conclusions} We have demonstrated the theoretical feasibility of the implementation of a DQC1 protocol in cold
atoms ensemble. The validity of the protocol extends to high  \textit{n} and allows to operate the DQC1 model to
compute sums over extremely large strings of numbers,  which make the computation classically intractable.  Quantum
computation has not yet been experimentally studied in  large Hilbert spaces, and the successful demonstration of  the
scalability of this protocol would be a major leap forward in the field.

The protocol presented in this work enables us to experimentally test the computational power of  quantum discord in a
regime never observed so far and it allows a thorough study  in high-dimension Hilbert spaces. In particular, by tuning
the purity of the control qubit, we can enter regimes with no entanglement and test the efficiency of the algorithm and
the power of discord as a resource for quantum computation.

It is important also to point out that non-trivial unitaries can be designed \cite{AddingControlTo2011x-qZhou} as part
of specific algorithms that would allow the implementation of a range of intractable tasks. We have proposed here a
general scheme that allows the implementation of a controlled-\textit{many-body} unitary. It is therefore possible to
envisage a new tool to explore the physics of many-body interacting systems, by exploiting DQC1 to enable the
measurement of the expectation values of operators acting on ensemble of strongly interacting qubits.

Besides providing a unique test for discord, this protocol can also be directly used for quantum phase estimation using
large ensembles \cite{PhysRevX.1.021022} and as a probe for quantum thermodynamics \cite{PhysRevLett.110.230601,
PhysRevLett.110.230602}. The proposed experiment can also be adapted to investigate quantum chaos
\cite{dPoulin2003TestingIntegrability, dPoulin2004ExponentialSpeedup} or to perform the overlap measurement scheme
\cite{c-sYu2012QuantumDissonanceIsRejected}, and has the potential to play a role in addressing some fundamental
questions like macrorealism \cite{amSouza2011AScatteringQuantumCircuit} and contextuality
\cite{oMoussa2010TestingContextuality}.

This work was  supported by EPSRC. The authors thank A. Datta, J. Bolton, K. Modi, H. Cable, C. MacCormick,  V. Vedral
 I. Beterov  for helpful discussions and W. Li and I. Lesanovsky for the C$_6$ coefficients.

\bibliography{bibdiscord}

\end{document}